\documentclass[usenatbib]{mn2e}
\usepackage{epsfig}
\usepackage{times}
\usepackage{lscape}
\usepackage{multirow}
\usepackage{multicol}
\usepackage{subfigure}
\usepackage{color}
\usepackage{amsfonts}
\usepackage{amssymb}

\begin{document}

\title[Simultaneous independent measurements of a truncated inner accretion disc] 
{Simultaneous independent measurements of a truncated inner accretion disc in the low/hard state of GX 399$-$4}
\author[D.~S. Plant, K.~O'Brien and R.~P. Fender] 
{D.~S. Plant$^{1,2}$, K.~O'Brien$^1$ and R.~P. Fender$^{1}$\\
$^1$Department of Physics, Astrophysics, University of Oxford, Keble Road, Oxford, OX1 3RH, UK\\
$^2$School of Physics and Astronomy, University of Southampton, Highfield, Southampton, SO17 1BJ, UK}
\maketitle


\begin{abstract}
We present results from three recent \emph{XMM-Newton} observations of GX 339$-$4 in the low/hard state, taken during the decay of a bright (peak $\sim\,0.05$\,L$_{\rm Edd}$) failed outburst. Uniquely, these are the first \emph{XMM-Newton} EPIC-pn observations of this source using an imaging mode, which significantly enhances the quality of the data at hand. In particular, thanks to the larger available bandpass, this allows an unprecedented constraint of the thermal accretion disc component, and the level of photoelectric absorption. We simultaneously measured the inner radius of the accretion disc via the broadened Fe K$\alpha$ line and the disc component. The two methods agree, and the measured radii show good consistency over the three epochs. We find that the inner radius is at 20--30\,$r_{\rm g}$, adding to the growing direct evidence for truncation of the inner accretion disc in the low/hard state.
\end{abstract}


\section{Introduction}\label{intro}

One of the most debated topics in the black hole X-ray binary (BHXRB) community is the accretion geometry in the canonical low/hard (hereafter `hard') state. The majority of BHXRBs exhibit bright (up to L$_{\rm X}$\,$\sim$\,L$_{\rm Edd}$) outbursts, which are separated by long periods of quiescence. During an outburst, a source transitions between the hard state, dominated by a hard power-law component peaking at $\sim100$\,keV, and the soft state, where the spectrum peaks around $\sim1$\,keV, and instead displays a quasi-blackbody shape. Outbursts \emph{always} begin and end in the hard state (see \citealt{Remillard06,Fender12} for recent reviews of BHXRB states). Some models predict that in quiescence ($\lesssim\,10^{-6}$\,L$_{\rm Edd}$) the accretion disc is substantially truncated $(\textgreater\,100$\,$r_{\rm g}$; \citealt{Narayan96,Esin97}), and in the soft state there is strong evidence that the accretion disc is at the innermost stable circular orbit (ISCO; \citealt{Gierlinski04, Steiner10, Dunn11}). It is likely, therefore, that the hard state represents a period of transition in the accretion geometry, but when, and how, this occurs, is of much debate.

The truncated disc model is based around the radiatively inefficient flow solutions (e.g. ADAFs; see \citealt{Yuan14} for a recent review), which are able to successfully explain accretion at very low $\dot{M}$. In this regime, the truncated disc gradually penetrates further into the hot flow as the accretion rate rises \citep{Remillard06,Done07}. This is able to explain a number of observables at increasing accretion rates; such as the softening of the power-law component, low reflection fractions, and the increase of characteristic frequencies in the power spectra \citep{VDK06,Done07, Plant14}. However, direct measurements of the inner disc radius have provided conflicting evidence as to whether the disc is truncated or not in the hard state.

The two leading spectroscopic methods to measure the disc inner radius come via the thermal accretion disc component and the broadened Fe K$\alpha$ line (see \citealt{McClintock11} and \citealt{Reynolds13} for the details of these two techniques). Attempts to determine the inner radius of the disc in this manner have led to conflicting results; some evidence suggests that the accretion disc may be at the ISCO at as low as $10^{-3}$\,L$_{\rm Edd}$ \citep{Rykoff07,Tomsick08,Reis10}, whereas other works, sometimes re-analysing the same data, suggest the accretion disc may truncated throughout the hard state (e.g. \citealt{Gierlinski08,Done10,Plant13,Kolehmainen13}). Key to resolving this debate is simultaneous measurements of the inner disc radius using the two methods. However, a recent study by \cite{Kolehmainen13}, using \emph{XMM-Newton} EPIC-pn timing mode observations of four BHXRBs, found the disc and reflection derived inner radii to be inconsistent.

In this letter, we simultaneously measure the disc and reflection inner radius using three recent \emph{XMM-Newton} observations of the black hole GX 339$-$4 taken in the EPIC-pn small window science mode. These data can potentially achieve a much better constraint on the accretion disc component due to the unprecedented bandpass calibration down to $\sim0.4$\,keV, thus offering the most stringent test yet of the truncated disc model. In \S2 we introduce the observations and the data reduction procedure, and then present our analysis in \S3. We find that both methods agree that the accretion disc is truncated, and the impact of this result is discussed in \S4.


\section{Observations and Data Reduction}\label{Sec:obs}

\begin{figure}
\centering
{\epsfig{file=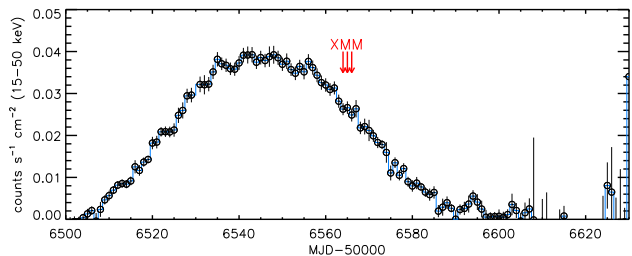, width=0.48\textwidth}}
\caption{A Swift-BAT (15--50\,keV) light-curve of the 2013 failed outburst of GX 339$-$4. The three \emph{XMM-Newton} observations used in this study are indicated, which caught the source after the peak of the outburst. The source remained in the hard state for the duration of the outburst.}
\label{Fig:BAT_lc}
\end{figure}

GX 339-4 was observed three times with \emph{XMM-Newton} over a period of three days (Table\,\ref{Tab:observations}). These observations took place during the decay of a failed outburst, which reached a peak X-ray luminosity of $\sim\,0.05$\,L$_{\rm Edd}$ (Fig.\,\ref{Fig:BAT_lc}). The EPIC-pn and EPIC-MOS cameras were operated in their small window modes, and the `thin' optical blocking filter was used. For this analysis we only used the EPIC-pn data, since the EPIC-MOS suffers significantly more from the effects of pile-up (by a factor of \textgreater\,5: XMM Users Handbook).

\begin{table}
\centering
\begin{tabular}{lcccc}\hline\hline
	&ObsID		& Date				& Exposure		& Count Rate	\\
	& 			& (UTC)				& (s)				& (counts s$^{-1}$)	\\\hline	
1	& 0692341201	& 2013-09-29 23:10:18	& 8 540			& 166 (59)	\\
2	& 0692341301	& 2013-09-30 22:50:37	& 9 426			& 162 (60)	\\
3	& 0692341401	& 2013-10-01 18:46:27 	& 15 036			& 154 (57)	\\\hline\hline
\end{tabular}
\caption{Observation log of the \emph{XMM-Newton} datasets taken in September-October 2013, which we analyse in this study. Net exposures correspond to that remaining after the reduction process and the live time of the science mode (71 per cent) is taken into account. Count rates in brackets refer to the count rate after the data have been corrected for pile-up.}
\label{Tab:observations}
\end{table}

Using the \emph{XMM-Newton} Science Analysis System (\textsc{sas}) version 13.5.0, the raw observation data files (ODFs) were reduced using the tool \textsc{epproc}, and the most recent current calibration files (CCFs) were applied. Spectra were extracted through \textsc{evselect}, using a circular region centred on the source with a radius of 45\,arcsec. Only single and double events (PATTERN$\leq$4) were used, and bad pixels were ignored (\#XMMEA\_EP and FLAG==0). Background regions were extracted using a circular region of the same size located away from the source. Using the standard tools, response and ancillary files were created (\textsc{rmfgen} and \textsc{arfgen}), and the spectra were required to have at least $20$ counts per channel using the FTOOL \textsc{grppha}. The nominal calibrated bandpass of the EPIC-pn small window mode is 0.3--10\,keV; however, we found there to be a strong residual in the pattern distributions created by \textsc{epatplot} below 0.4\,keV for all three observations. This was observed as a dip/peak in the single/double distributions, and has been observed a number of times in small window mode datasets. We expect the origin of these features to be from background noise, and they were greatly reduced, but still present, when a background event file was included (\texttt{withbackgroundset=yes} in \textsc{epatplot}). The features were strongest in Observation 3, which occurred at the start of orbit 2530, thus it is likely that they are due to background radiation. We therefore applied a lower limit of 0.4\,keV for all of the spectra, and we also ignored the 1.75--2.35\,keV region where strong features were observed. These are often seen when observing bright sources, and are likely to be of instrumental origin.

\renewcommand{\arraystretch}{1.3}
\begin{table}
\centering
\begin{tabular}{lccc}\hline \hline
Parameter						& Model--1				& Model--2				& Model--2b				\\\hline\hline
\multicolumn{4}{c}{Observation 1} \\ \hline
N$_{\rm H}$ (10$^{22}$ cm$^{-2}$)		& 0.50\,$\pm\,{0.01}$		& 0.64\,$\pm\,{0.02}$		& 0.64\,$\pm\,{0.02}$		\\
A$_{\rm O}$						& 						&						& 1.34\,$\pm\,{0.06}$		\\
T$_{\rm in}$ (keV)					&						& 0.20\,$\pm\,{0.01}$		& 0.19\,$\pm\,{0.01}$		\\
N$_{\rm DBB}$ (10$^3$)				& 						& 10.76$^{+4.86}_{-3.53}$	& 24.83$^{+9.40}_{-7.21}$	\\
$\Gamma$						& 1.62\,$\pm\,{0.01}$		& 1.62\,$\pm\,{0.01}$		& 1.63\,$\pm\,{0.01}$		\\
N$_{\rm Comp}$					& 0.16\,$\pm\,{0.01}$		& 0.16\,$\pm\,{0.01}$		& 0.17\,$\pm\,{0.01}$		\\ \hline
$\chi{^2}/\nu$						& 2233/1705				& 2026/1703				& 1943/1702				\\ \hline\hline
\multicolumn{4}{c}{Observation 2} \\\hline
N$_{\rm H}$ (10$^{22}$ cm$^{-2}$)		& 0.50\,$\pm\,{0.01}$		& 0.61\,$\pm\,{0.02}$		& 0.61$^{+0.02}_{-0.01}$		\\
A$_{\rm O}$						&						&						& 1.30$\pm{0.06}$			\\
T$_{\rm in}$ (keV)					&						& 0.22\,$\pm\,{0.02}$		& 0.20\,$\pm\,{0.01}$		\\
N$_{\rm DBB}$ (10$^3$)				&						& 5.77$^{+2.67}_{-1.92}$		& 13.32$^{+5.37}_{-4.02}$	\\
$\Gamma$						& 1.62\,$\pm\,{0.01}$		& 1.60\,$\pm\,{0.01}$		& 1.61\,$\pm\,{0.01}$		\\
N$_{\rm Comp}$					& 0.16$\pm{0.01}$			& 0.16$\pm{0.01}$			& 0.16$\pm{0.01}$			\\ \hline
$\chi{^2}/\nu$						& 2379/1743				& 2173/1741				& 2106/1740				\\ \hline \hline
\multicolumn{4}{c}{Observation 3} \\ \hline
N$_{\rm H}$ (10$^{22}$ cm$^{-2}$)		& 0.49$\pm{0.01}$			& 0.61$\pm{0.02}$			& 0.61$\pm{0.02}$			\\
A$_{\rm O}$						&						& 						& 1.34$\pm{0.05}$			\\
T$_{\rm in}$ (keV)					& 						& 0.22$\pm{0.01}$			& 0.20$\pm{0.01}$			\\
N$_{\rm DBB}$ (10$^3$)				& 						& 5.56$^{+1.91}_{-1.48}$		& 13.05$^{+3.14}_{-3.97}$	\\
$\Gamma$						& 1.61$\pm{0.01}$			& 1.59$\pm{0.01}$			& 1.60$\pm{0.01}$			\\
N$_{\rm Comp}$					& 0.15$\pm{0.01}$			& 0.15$\pm{0.01}$			& 0.15$\pm{0.01}$			\\ \hline
$\chi{^2}/\nu$						& 2610/1790				& 2220/1788				& 2092/1787				\\\hline \hline
\end{tabular}
\caption[Results from various continuum model fits]{Results from various continuum model fits.\newline
Model--1: \textsc{tbnew\_feo$\ast$(nthcomp)}\newline
Model--2: \textsc{tbnew\_feo$\ast$(diskbb+nthcomp)}\newline
Model--2b: The same as Model--2, but the O abundance in \textsc{tbnew\_feo} (A$_{\rm O}$) is now a free parameter.}
\label{Tab:continuum}
\end{table}

We applied the \textsc{sas} tool \textsc{epatplot} to all of the datasets to check for photon and pattern pile-up, which was found to be high in all three observations. To mitigate the pile-up we investigated a number of annulus extraction regions for the spectra, and found an inner cavity with a radius of 11.5 arcsec to be very acceptable. This reduced the estimated level of pile-up to to be less than 1 per cent for both the single and double events between 0.4--10\,keV. All of the analysis in this letter was performed using \textsc{xspec} v.12.8.0, and all quoted errors are at the 90 per cent confidence level. We use the `wilm' abundances and `vern' cross-sections for all of the analysis.

\section{Analysis and Results}\label{Sec:analysis}

\begin{figure}
\centering
{\epsfig{file=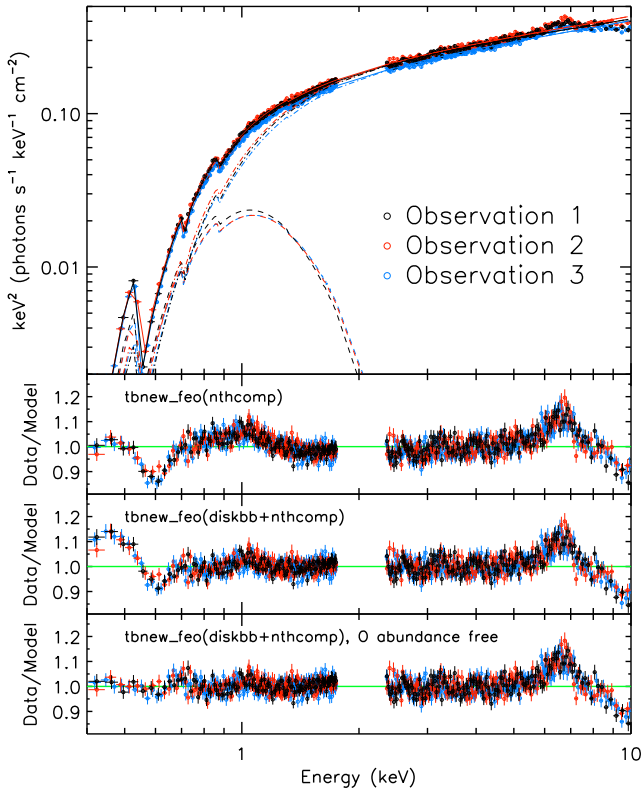, width=0.48\textwidth}}
\caption[Unfolded spectra and model residuals to the best-fit continuum model]{Unfolded spectra and model residuals to the best-fit continuum model (Model--2b; Table\,\ref{Tab:continuum}). There is a clear excess and edge around 7\,keV, which strongly suggests an Fe emission line corresponding edge is present in the spectra. Solid, dash and dot-dash lines describe the total, \textsc{diskbb} and \textsc{nthcomp} model components respectively.}
\label{Fig:fit1}
\end{figure}

The hard state X-ray spectrum of BHXRBs is often successfully modelled with a power-law, which is most commonly associated with inverse Comptonisation of photons in a corona of hot ($\sim100$\,keV) electrons. Therefore, to begin, we fit each spectrum with an absorbed power-law (\textsc{tbnew\_feo$\ast$powerlaw} in XSPEC; \citealt{Wilms00}). The fit was reasonable ($\chi^2_{\nu}/\nu=1.43/5238$; Table\,\ref{Tab:continuum}), and both the photon index and column density were typical for GX 339$-$4 in the hard state (e.g. \citealt{Kolehmainen13}).

However, a power-law is not a physical model, and can lead to an excess of photons a low energies. Assuming that the hard power-law tail arises through Comptonisation of disc photons, there should be a low-energy cut-off associated with the disc temperature. Thermal Comptonisation is the favoured interpretation for the hard state power-law (see \citealt{Done07} for a detailed review), thus we replaced the \textsc{powerlaw} model with \textsc{nthcomp} \citep{Zdziarski96}. This model includes a high energy cutoff, parameterised by the electron temperature kT$_{\rm e}$, which we fixed to be 50\,keV \citep{Done07}. Because of the soft bandpass used in this study, this parameter could not be constrained, but the cut-off is not expected to be low enough to affect the spectrum \textless10\,keV. This model also includes a low energy rollover, defined by a seed photon temperature kT$_{\rm bb}$, which we fixed to be 0.2\,keV (a typical value for the hard state; \citealt{Reis10,Kolehmainen13}). To this end, we also set the seed spectrum to be a disc-blackbody. The other free parameters were the photon index and normalisation, like for the \textsc{powerlaw} model. Using \textsc{nthcomp} the fit was slightly improved ($\chi^2_{\nu}/\nu$=1.38/5238; Table\,\ref{Tab:continuum}), but not significantly.

A strong excess at $\sim\,$7\,keV, and drop above 8.5\,keV, were clear (Fig. \ref{Fig:fit1}), suggesting that a significant amount of reflection was present, which is ubiquitously observed in the hard state of GX 339$-$4 (see e.g. \citealt{Dunn08,Plant14}). Also, an excess peaking at $\sim\,1$\,keV, and a strong edge at 0.6\,keV, were further clear residuals in the spectra. The excess at $\sim\,1$\,keV is likely to be due to thermal emission from the accretion disc, which is often observed from hard state sources with soft X-ray instruments \citep{Miller06,Reis10,Kolehmainen13}. We therefore added a multi-colour disc-blackbody model (\textsc{diskbb}; \citealt{Mitsuda84}), which considerably improved the fit ($\chi^2_{\nu}/\nu$=1.22/5232; Table\,\ref{Tab:continuum}). The disc component contributes $\sim\,17$ per cent of the total unabsorbed 0.4--10\,keV flux.

However, the fit was still poor \textless\,0.7\,keV (Fig. \ref{Fig:fit1}), the main cause of which was a strong edge-like feature at $\sim0.6$\,keV, coincident with the O K edge at 0.54\,keV (see e.g. \citealt{Wilms00}). We therefore allowed the O abundance to be a free parameter, which further improved the fit ($\chi^2_{\nu}/\nu$=1.17/5229; Table\,\ref{Tab:continuum}). All of the observations settled on an O abundance of $\sim\,1.34$, and meant that the soft bandpass was now excellently fit ($\chi^2_{\nu}$=1.04 below 5\,keV).

As can be seen in Fig.\,\ref{Fig:fit1}, a strong Fe K$\alpha$ emission line and corresponding edge were still present in the spectra. To model this we added the X-ray reflection model \textsc{relxill} \citep{Garcia14}, which combines the relativistic \textsc{relconv} code of \cite{Dauser10} with the angle-resolved X-ray reflection table \textsc{xillver} of \cite{Garcia13}. This represents the first code to prescribe the correct reflection spectrum for each relativistically calculated emission angle, as opposed to the usual relativistic convolution treatment to the spectrum (e.g. \textsc{relconv$\ast$xillver}).

\renewcommand{\arraystretch}{1.3}
\begin{table}
\centering
\begin{tabular}{lccc}\hline \hline
Parameter						& Observation 1		& Observation 2			& Observation 3					\\ \hline
N$_{\rm H}$ (10$^{22}$ cm$^{-2}$)	& 0.76$^{+0.01}_{-0.02}$		& 0.74$\pm{0.01}$			& 0.74$^{+0.03}_{-0.02}$				\\
A$_{\rm O}$					& 1.52$^{+0.03}_{-0.06}$		& 1.50$^{+0.05}_{-0.04}$		& 1.55$^{+0.04}_{-0.02}$				\\
T$_{\rm in}$ (keV)				& 0.16$\pm{0.01}$			& 0.16$\pm{0.01}$			& 0.17$\pm{0.01}$					\\
N$_{\rm DBB}$ (10$^5$)			& 1.34$^{+0.28}_{-0.18}$		& 1.09$^{+0.21}_{-0.19}$		& 0.89$^{+0.43}_{-0.20}$				\\
$r_{\rm in\,(DBB)}$ ($r_{\rm g}$)		& 25$^{+3}_{-2}	$		& 29$\pm{3}$ 				& 26$^{+3}_{-6}$					\\
$\Gamma$					& 1.62$\pm{0.02}$			& 1.65$\pm{0.01}$			& 1.59$^{+0.02}_{-0.03}$				\\
N$_{\rm Comp}$				& 0.092$\pm{0.01}$			& 0.14$\pm{0.01}$			& 0.08$\pm{0.01}$					\\
$r_{\rm in\,(Fe)}$ ($r_{\rm g}$)		& 21$^{+17}_{-9}$			& 27$^{+6}_{-6}$ 			& 16$^{+7}_{-4}$					\\
$i (^{\circ})$					& \multicolumn{3}{c}{30$^{+5}_{-4}$}													\\ 
$\log{\xi}$ (erg\,cm\,s$^{-1}$)		& 3.17$\pm{0.06}$			& 3.03$^{+0.05}_{-0.03}$		& 3.17$^{+0.06}_{-0.05}$				\\
N$_{\rm Fe}$ (10$^{-6}$)		& 0.34$\pm{0.03}$			& 0.26$^{+0.03}_{-0.04}$		& 0.32$^{+0.03}_{-0.02}$				\\ \hline
$\chi{^2}/\nu$					& \multicolumn{3}{c}{5322/5219}													\\ \hline \hline
\end{tabular}
\caption{The best-fit model and MCMC derived 90 per cent confidence limits for fits to the three GX 339$-$4 observations. The inclination parameter is fitted jointly between the three spectra. `DBB' and `Fe' refer to the \textsc{diskbb} and \textsc{relxill} components respectively.}
\label{Tab:reflection}
\end{table}
\begin{figure}
\centering
{\epsfig{file=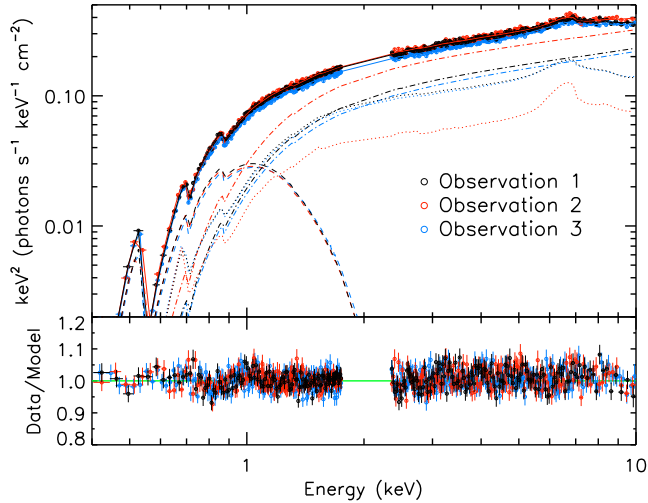, width=0.48\textwidth}}
\caption[Unfolded spectra and model residuals to the best-fit continuum+reflection model]{Unfolded spectra and model residuals to the best-fit continuum+reflection model (Table\,\ref{Tab:reflection}), which represents an excellent fit over the entire bandpass. Solid, dash, dot-dash and dotted lines describe the total, \textsc{diskbb}, \textsc{nthcomp} and \textsc{relxill} model components respectively.}
\label{Fig:fit2}
\end{figure}
\begin{figure}
\centering
{\epsfig{file=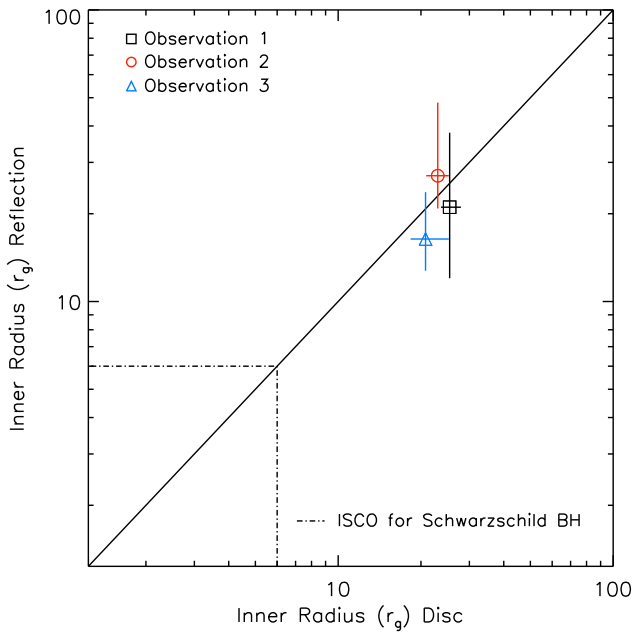, width=0.48\textwidth}}
\caption{The inner radius of the accretion disc, fitted by the disc (x-axis) and reflection (y-axis) components (Table \ref{Tab:reflection}). The dot-dash line indicates the innermost stable circular orbit for a zero spin black hole, showing that both methods, from all three observations, are consistent with a truncated accretion disc.}
\label{Fig:radii}
\end{figure}

We fixed the spin to be 0.9\footnote{A spin of 0.9 is a compromise between previous conflicting spin measurements of GX 339-4 \citep{Reis08,Kolehmainen10}. We tested different values (0, 0.5, 0.998) and found them to not affect the results, which should be expected since the spin has little effect on the Fe profile for the values of $r_{\rm in}$ we measure \citep{Dauser10}.}, and assumed an emissivity profile of $R^{-3}$. We also linked the photon index in \textsc{nthcomp} to the illuminating index in \textsc{relxill}. The ionisation parameter $\xi$, the disc inner radius $r_{\rm in}$, and the model normalisation were all fitted freely. Finally, the inclination was fitted jointly by the three spectra. For all three observations the inner radius was found to be moderately truncated at $\sim\,20\,r_{\rm g}$, and the fit was excellent over the whole bandpass ($\chi^2_{\nu}$=1.02/5219; Fig.\,\ref{Fig:fit2}). Given the excellent constraint on the disc emission allowed by the 0.4--10\,keV bandpass, we were able to compare how the inner radius derived from the disc component compares to that from the reflection. The normalisation of \textsc{diskbb} is defined as \begin{equation}
N_{\rm BB}=1.19\left(\frac{r_{\rm in}}{D_{\rm 10\,{kpc}}}\right)^2\cos{i}\label{eq1}
\end{equation} where $D$ is the distance to the source in units of 10\,kpc, $r_{\rm in}$ is the disc inner radius in km, and $i$ is the inclination of the disc, which we assumed to be the same as fitted by the reflection component (Table\,\ref{Tab:reflection}). The 1.19 factor accounts for the spectral hardening factor $f_{\rm col}$ and that $T_{\rm in}$ occurs at a radius somewhat larger than $r_{\rm in}$ \citep{Kubota98}. We find that the inner radii derived from the disc and reflection are very consistent. To probe this connection further we performed a Markov Chain Monte Carlo (MCMC) analysis on the best-fit parameters. Six 55,000 element chains were produced, of which the first 5000 elements were discarded (`burnt'). Each chain was run from a random perturbation away from the best fit, and the chain proposal was taken from the diagonal of the subsequent covariance matrix. The probability distributions were assumed to be Gaussian, and a rescaling factor of $5\times10^{-4}$ was applied. The confidence limits derived from the resultant 300,000 element chain are presented in Table\,\ref{Tab:reflection}.

The agreement between the inner radii calculated from the reflection and disc components is very good, despite the relatively short exposure of the data. Figure \ref{Fig:radii} compares the respective radii and presents compelling evidence that the inner disc is truncated at 20--30\,$r_{\rm g}$. Furthermore, the disc component radii are extremely consistent between the three observations, despite the weakness of the component in the hard state. The confidence limits of the reflection radii are larger, but still good considering the relatively short exposures and low flux of the system.

Recent works have shown that the emissivity profile is complex close to the black hole (\textless\,10\,$r_{\rm g}$; \citealt{Wilkins12,Dauser13}). We therefore relaxed the assumption that the emissivity profile is R$^{-3}$; however, rather than fitting a broken power-law emissivity, which is not at all physical, we instead switched to an additional version of \textsc{relxill} called \textsc{relxilllp}. This model fits the height of the illuminating X-ray source, applying the corresponding emissivity profile to the reflection spectrum \citep{Dauser13,Garcia14}. This model can also self-consistently fit the power-law and reflection components, ensuring that the correct reflection fraction is computed as a function of source height and disc inner radius. We therefore removed \textsc{nthcomp} and allowed \textsc{relxilllp} to self-consistently fit the two components. We find that the results are very consistent with those assuming R$^{-3}$, and the fit is improved ($\Delta\chi^2=-21$; Table \ref{Tab:reflection_lp}). In particular, the disc inner radius fitted from both the disc and reflection components is almost unchanged. The fitted source height $h$ is consistent within errors over the three observations, and is measured to be at 14--70\,$r_{\rm g}$ (90 per cent confidence). The constraint on $h$ is poor since the emissivity profiles of a broad range of source heights converge at the values of $r_{\rm in}$ being fitted \citep{Dauser13}. Much better constraints on $h$ should be possible at higher accretion rates when the disc is expected to be less truncated.


\section{Discussion and Conclusions}

In this letter, we have presented analysis of three recent observations of GX 339$-$4 in the hard state during the decay phase of a failed outburst. These observations utilised the small window science mode, and represent one of the first hard state observations of a black hole using an \emph{XMM-Newton} EPIC-pn imaging mode. This presents a well calibrated bandpass down to 0.4\,keV, whereas the fast modes that are usually employed are restricted to above 0.7\,keV. The result is an unprecedented view of the accretion disc component, which typically peaks at $\sim\,0.2$\,keV in the hard state. This also removes the calibration uncertainties associated with the fast modes; such as reliable pile-up correction (see \citealt{Miller10}), the lack of source-free background regions \citep{Ng10}, and issues with charge-transfer inefficiency at high count rates \citep{Walton12}.

Aided by the excellent constraint of the disc component we were able to measure the inner radius from both the disc and reflection components, which both agree on a moderately truncated inner disc. The jointly fit inclination is, however, smaller than expected, given the mass function constraint of \cite{Munoz08}. As shown in \cite{Plant13}, a higher fitted inclination would act to increase the reflection inner radius, since the larger doppler shifts at higher inclination broaden the Fe profile. We performed fits with larger fixed values of $i$, which confirmed that the fitted reflection inner radius does indeed increase. The fitted radii are considerably smaller than those found in the hard state of GX 339$-$4 by \cite{Plant13} and \cite{Kolehmainen13}; however, this is, at least in part, due to the smaller inclination. From Eq.\,\ref{eq1}, it is clear that larger $i$ would have the same effect on the inner radius measured by the disc component. 
\renewcommand{\arraystretch}{1.3}
\begin{table}
\centering
\begin{tabular}{lccc}\hline \hline
Parameter					& Observation 1		& Observation 2			& Observation 3					\\ \hline
$h$ ($r_{\rm g}$)				& 37$^{+33}_{-13}$		& 17$^{+5}_{-3}$ 				& 26$^{+18}_{-7}$					\\
$r_{\rm in\,(DBB)}$ ($r_{\rm g}$)		& 25$\pm{2}$			& 29$\pm{3}$ 				& 26$^{+3}_{-1}$				\\
$r_{\rm in\,(Fe)}$ ($r_{\rm g}$)		& 20$^{+4}_{-5}$		& 29$\pm{2}$ 		& 16$\pm{2}$				\\
$i (^{\circ})$					& \multicolumn{3}{c}{33$\pm{3}$}												\\ \hline
$\chi{^2}/\nu$					& \multicolumn{3}{c}{5301/5219}													\\ \hline \hline
\end{tabular}
\caption{The best-fit model and MCMC derived 90 per cent confidence limits for the fit with \textsc{relxilllp}. All of the parameters not listed in this table were consistent within errors with those in Table \ref{Tab:reflection}. The inclination parameter was again fitted jointly between the three spectra.}
\label{Tab:reflection_lp}
\end{table}

To conclude, this is the first time that the disc and reflection methods have been able to simultaneously test truncation of the inner accretion disc in the hard state. This adds to recent direct evidence for truncation in the hard state, which together are finally establishing certainty in the hard state accretion geometry.

\section*{Acknowledgements}
DSP acknowledges financial support from the STFC. This project was funded in part by European Research Council Advanced Grant 267697 4-pi-sky: Extreme Astrophysics with Revolutionary Radio Telescopes. This letter is based on observations obtained with \emph{XMM-Newton}, an ESA science mission with instruments and contributions directly funded by ESA Member States and the USA (NASA).
\bibliographystyle{mn2e_fix}
\bibliography{refs}

\end{document}